\documentclass[twocolumn,tighten]{aastex62}

\usepackage{gensymb}
\usepackage{amsmath}

\newcommand{\halpha}{H$\alpha$}
\newcommand{\hbeta}{H$\beta$}

\newcommand{\hi}{\ion{H}{1}}

\newcommand{\henir}{{\ion{He}{1}}$~\lambda$10830}

\newcommand{\mdot}{$\dot{\text{M}}$}
\newcommand{\wten}{W$_{10}$}
\def\msun{\rm M_{\sun}}
\def\rsun{\rm R_{\sun}}

\def\msunyr{\rm{M_{\sun} \, yr^{-1}}}
\def\kms{\rm \, km \, s^{-1}}

\newcommand{\tmax}{$T_{\rm max}$}

\received{---- --, ----}
\revised{---- --, ----}
\accepted{---- --, ----}
\submitjournal{ApJ}

\shorttitle{Complex Magnetospheric Flows in CVSO 1335}
\shortauthors{Thanathibodee et al.}

\begin{document}

\title{Complex Magnetospheric Accretion Flows in Low Accretor CVSO 1335}

\author[0000-0003-4507-1710]{Thanawuth Thanathibodee}
\affiliation{Department of Astronomy, University of Michigan, 323 West Hall, 1085 South University Avenue, Ann Arbor, MI 48109, USA}
\email{thanathi@umich.edu}

\author[0000-0002-3950-5386]{Nuria Calvet}
\affiliation{Department of Astronomy, University of Michigan, 323 West Hall, 1085 South University Avenue, Ann Arbor, MI 48109, USA}

\author{James Muzerolle}
\affiliation{Space Telescope Science Institute, 3700 San Martin Drive, Baltimore, MD 21218, USA}

\author[0000-0001-7124-4094]{C\'esar Brice\~no}
\affiliation{Cerro Tololo Inter-American Observatory, National Optical Astronomy Observatory, Casilla 603, La Serena, Chile}

\author[0000-0002-1650-3740]{Ramiro Franco Hern\'andez}
\affiliation{Instituto de Astronom\'ia y Meteorolog\'ia, Universidad de Guadalajara, Avenida Vallarta No. 2602, Col. Arcos Vallarta, CP 44130, Guadalajara, Jalisco, M\'exico}

\author[0000-0001-8284-4343]{Karina Mauc\'o}
\affiliation{N\'ucleo Milenio Formaci\'on Planetaria - NPF, Universidad de Valpara\'iso, Av. Gran Breta\~na 1111, Valpara\'iso, Chile}
\affiliation{Instituto de F\'isica y Astronom\'ia, Facultad de Ciencias, Universidad de Valpara\'iso, Av. Gran Breta\~na 1111, Valpara\'iso, Chile}

\begin{abstract}
Although the magnetospheric accretion model has been extensively applied to T Tauri Stars with typical mass accretion rates, the very low accretion regime is still not fully explored. Here we report multi-epoch observations and modeling of CVSO~1335, a 5 Myr old solar mass star which is accreting mass from the disk, as evidenced by redshifted absorption in the H$\alpha$ profile, but with very uncertain estimates of mass accretion rate using traditional calibrators. We use the accretion shock model to constraint the mass accretion rate from the Balmer jump excess measured with respect to a non-accreting template, and we model the H$\alpha$ profile, observed simultaneously, using magnetospheric accretion models. Using data taken on consecutive nights, we found that the accretion rate of the star is low, $4-9 \times 10^{-10} \,$ M$_{\odot}$\,yr$^{-1}$, suggesting a variability on a timescale of days. The observed H$\alpha$ profiles point to two geometrically isolated accretion flows, suggesting a complex infall geometry. The systems of redshifted absorptions observed are consistent with the star being a dipper, although multi-band photometric monitoring is needed to confirm this hypothesis. 
\end{abstract}


\section{Introduction} \label{sec:intro}

Low-mass pre-main sequence stars, or T Tauri stars (TTS), are formed surrounded by disks and evolve accreting mass from these disks.
The accretion of the material from the inner region of protoplanetary disk onto the star follows the magnetospheric accretion paradigm \citep{hartmann2016}. 
Under this framework, gaseous material, heated by the stellar radiation field and other processes, flows along the magnetic field lines onto the star, creating an accretion shock at the base of the flow. 
The emission from the accretion shock is observed as an excess over the stellar photosphere \citep{calvet1998}.
Emission lines form in the magnetospheric accretion flows, so that the kinematics of the flow can be inferred from the line profiles.
Numerical simulations of magnetized stars \citep[e.g.][]{romanova2003} as well as magnetospheric accretion models \citep{hartmann1994,muzerolle1998a,muzerolle2001,kurosawa2011} have confirmed this picture for accreting T Tauri stars (Classical T Tauri star; CTTS) and have provided insight into the physical properties of accretion, including the geometry of the flows.

One of the most important properties of accretion is the mass accretion rate {\mdot}. This can be estimated by either directly measuring the excess over the photosphere and inferring the accretion luminosity, 
L$_{\rm acc} = {\rm GM}_{\star}\dot{\rm M}/{\rm R}_{\star}$, 
or by using emission lines. 
Specifically, the excess flux over the photosphere can be extracted from the optical flux by measuring the veiling of photospheric absorption lines and measured directly in the UV \citep[e.g.][]{ingleby2013}.
The accretion luminosity can then be measured using accretion shock models \citep{calvet1998,robinson2019}, 
or slab models \citep[e.g.][]{gullbring1998,herczeg2008,manara2016,alcala2017},
to fit the excess and account for the flux outside the wavelength regions where the excess is observed.
Emission lines have been used to infer the mass accretion rate via empirical relationships between the line luminosity or line width 
and the accretion luminosity \citep{muzerolle1998b,natta2004,calvet2004,ingleby2013,alcala2014}.
These relationships have been calibrated with direct measurements of accretion luminosity from the excess over the photosphere, obtained simultaneously in many cases.
A more direct method to measure accretion rates from emission lines is by modeling the line profiles with magnetospheric accretion models. 
This has been done for hydrogen Balmer lines \citep{muzerolle2001,natta2004,espaillat2008}, Na~D line \citep{muzerolle2001}, and {\henir} line \citep{fischer2008,kurosawa2011}. Modeling emission lines also provides information about accretion geometry.

Finding the method that best estimates the  mass accretion rates is especially relevant for the stars with the lowest accretion rates, the {\it low accretors}. Population studies show that 
the
mass accretion rate scales with stellar mass \citep[e.g.][]{muzerolle2003a,herczeg2008,alcala2014,manara2015}, suggesting that a mass range is implied when a star is designated as a low accretor. For example, the current detection limit of mass accretion for solar-mass star is $\sim10^{-10}\,\msunyr$ \citep{ingleby2013,manara2013}, whereas this accretion rate is not unusual for mid-M type stars \citep[e.g.][]{alcala2017}.

As expected from viscous evolution of protoplanetary disks, the mass accretion rate onto the star decreases with time \citep{hartmann1998}. 
Studies of many star-forming regions also show that the frequency of accretors, as well as the frequency of disk-bearing T Tauri stars, in a given population decreases as the age of the population increases \citep{fedele2010,hernandez2008,briceno2019}. 
However, it is unclear how accretion proceeds at very low accretion rates and how it finally stops. 
To understand processes occurring at the last stages of accretion it is necessary to carry out systematic studies of T Tauri stars accreting at very low accretion rates.
This is the main motivation for our ongoing observational and modeling program to search and characterize low accretors.

As the initial result of our study of low accretors, we presented the characterization of the inner disk of three 5\,Myr T Tauri stars in \citet{thanathibodee2018}. 
One target, CVSO~1335, was particularly interesting. The star is a pre-main sequence solar analog with M$_{\star}=0.87\,\msun$, R$_{\star}=1.58\,\rsun$, and spectral type of K5. 
The star, located in the 5 Myr old Ori~OB1b subassociation \citep{briceno2019}, is a CTTS, based on the presence of redshifted absorption in the {\henir} line. 
\citet{thanathibodee2018} showed that the protoplanetary disk surrounding the star is gas-rich, as indicated by fuorescent FUV H$_2$ emission, while the spectral energy distribution indicates that it is a transitional disk with a gap depleted of small dust. 
That work also showed that the star had complex {\halpha} line profiles in several epochs, which complicated the measurement of its mass accretion rate. 
The accretion rates determined with different indicators differed by more than 3 orders of magnitude. 
The lower range of the measured mass accretion rate at $\sim10^{-10}\,\msunyr$ would suggest that the star is a low accretor for its mass.
The disagreement between the accretion diagnostics and the complex features in {\halpha}, especially the persistent low-velocity redshifted absorption which is found in AA~Tau-like stars \citep[e.g.][]{bouvier2007,fonseca2014} and other dippers \citep[e.g.][]{alencar2018}, makes this star an ideal target for a detailed study of accretion properties in low accretors. 

Here we report the observations and the characterization of the accretion rate and the accretion geometry in the low accretor CVSO~1335. Optical spectra of the star are presented in \S\ref{sec:observation}, with the analysis and modeling in \S\ref{sec:result}. The implication of the model results are presented in \S\ref{sec:discussion} with a summary in \S\ref{sec:summary}.

\section{Observations} \label{sec:observation}
The analysis of the {\halpha} line in \citet{thanathibodee2018}
was based on observations with the moderate-resolution (R$\sim$4100, 73\,$\kms$) MagE Spectrograph on the 6.4m Magellan Baade telescope at the Las Campanas Observatory in Chile, as well as the high-resolution (R$\sim$14000, 22\,$\kms$) Goodman Spectrograph on the CTIO SOAR Telescope.
The MagE spectrograph covers the full optical range from 3200-8200{\AA}, giving access to the Balmer jump.

We report here additional spectra observed on 2017 November 27 and 28, which we reduced similarly to the MagE spectra reported in \citet{thanathibodee2018}.
In total, we have four MagE spectra observed in consecutive nights, and one Goodman spectra observed about two months earlier.
Table \ref{tab:obs} shows the details of all observations of CVSO~1335.

\begin{deluxetable}{llccc}[t]
\tablecaption{Summary of Observations \label{tab:obs}}
\tablehead{
\colhead{Instrument} 	&
\colhead{Start Date} &
\colhead{Exp. time} &
\colhead{Airmass} &
\colhead{SNR\tablenotemark{a}} \\
\colhead{} 		&
\colhead{(UT)} 	&
\colhead{(sec)} &
\colhead{} &
\colhead{}
}
\startdata
Goodman 	& 2017 Sep 18\tablenotemark{b}	& $3 \times 600$ & 1.22 & 35 \\
MagE    	& 2017 Nov 27	& $3 \times 600$ & 1.75 & 290 \\
     	& 2017 Nov 28	& $3 \times 600$ & 1.14 & 260 \\
     	& 2017 Nov 29\tablenotemark{b}	& $2 \times 900$ & 1.14 & 280 \\
		& 2017 Nov 30\tablenotemark{b}& 600+900 		 & 1.14 & 260 \\
\enddata
\tablenotetext{a}{Signal-to-noise at 6560\,{\AA}.}
\tablenotetext{b}{Reported in \citet{thanathibodee2018}.}
\end{deluxetable}

\begin{figure*}[t]
\epsscale{1.18}
\plotone{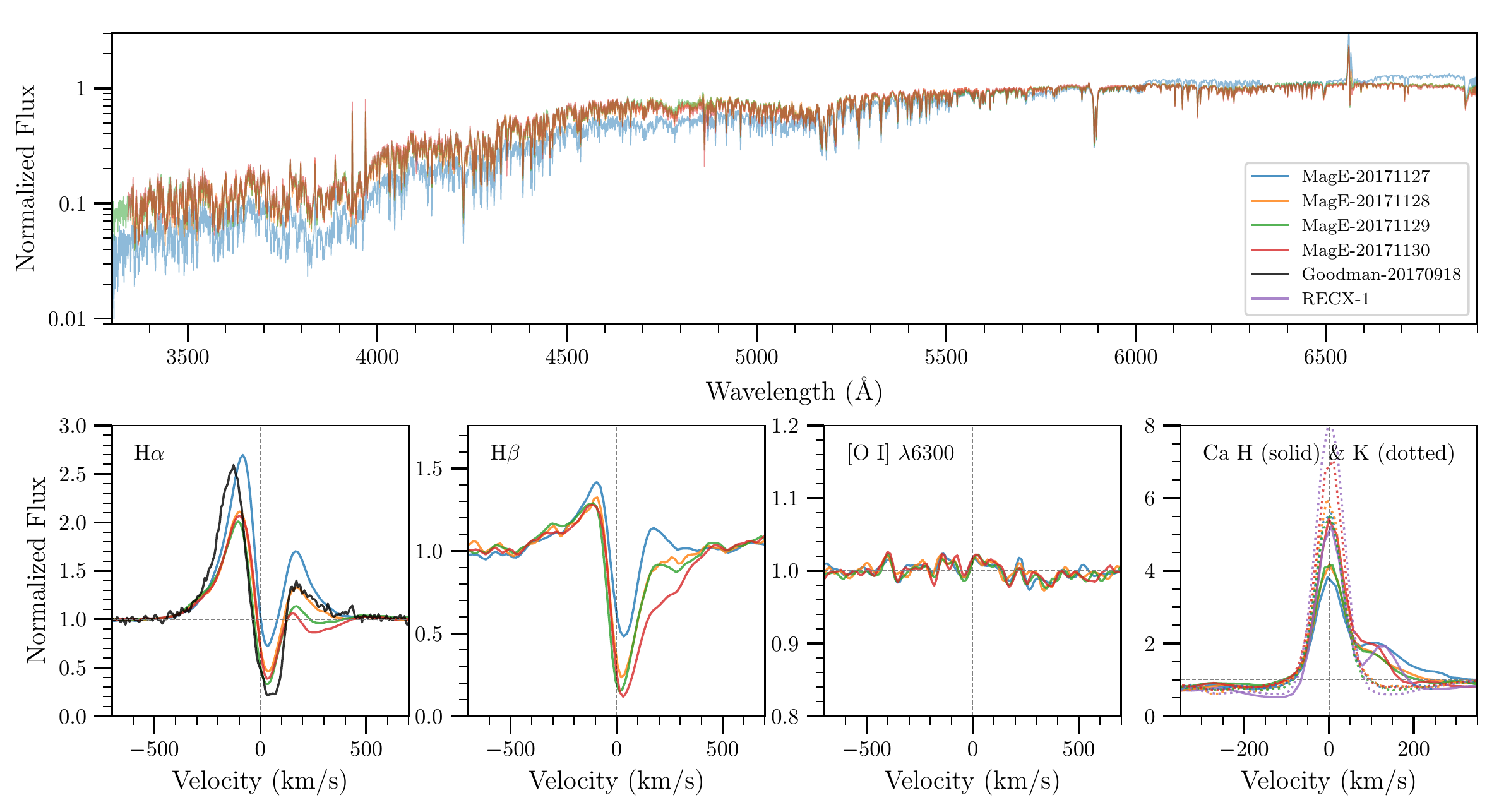}
\caption{\emph{Top:} 
Optical spectra of CVSO~1335 observed with the MagE spectrograph (R$\sim$4100). 
The spectra are normalized to unity at $\sim$6000\AA. 
\emph{Bottom:} Profiles of the {\halpha}, {\hbeta}, \ion{O}{1} lines, and the \ion{Ca}{2}~HK doublet. We also include a high resolution {\halpha} profile observed by the Goodman spectrograph (R$\sim$14000). For the hydrogen lines, the chromospheric contribution have been subtracted by the convolved spectrum of RECX~1, a non-accretor observed by UVES. The hydrogen Balmer lines shows complex and variable features, in particular
a low-velocity redshifted absorption. This absorption is especially strong for the {\hbeta} line observed on the last 3 epochs. The \ion{O}{1}\,$\lambda6300$ lines show no detection. The \ion{Ca}{2} lines are symmetric, showing minimal variation
in epochs separated by 1 day, and increases in strength on 20171130. The emission to the right of the \ion{Ca}{2}~H line is from H$\epsilon$. The \ion{Ca}{2} lines of RECX~1 are comparable to those of CVSO~1335 observed on 20171130.}
\label{fig:observation}
\end{figure*}

The top panel of Figure~\ref{fig:observation} shows the optical spectra of CVSO~1335 observed with the MagE spectrograph. 
Several emission lines are presented in the spectra, including the hydrogen Balmer lines and the Ca H \& K doublet. 
We do not detect the \ion{O}{1}\,$\lambda6300$ in the MagE spectra, suggesting that the stellar or disk wind is very weak in the system, and the contrast between the line and the continuum could be low for a K5 star. 
The lower panel in Figure~\ref{fig:observation} shows the corresponding {\halpha} profiles from the MagE and Goodman spectra, 
as well as other emission lines. To remove the photospheric contribution from the \ion{O}{1} lines and the chromospheric and photospheric contributions from the {\hi} lines, which are generally found in active young stars \citep{manara2013,manara2017},  we subtracted the line profile of RECX~1, a standard non-accretor (Weak T Tauri star; WTTS), from the CVSO~1335 profiles. The standard star has similar stellar parameters as those of CVSO~1335 (SpT=K5-6, M$_{\star}=0.9\,\msun$, R$_{\star}=1.8\,\rsun$, age$\sim$5-9\,Myr), so its spectrum can be used to represent the underlining photospheric and chromospheric emission of the target \citep{ingleby2011b}.
The spectrum of RECX~1 was obtained by the UVES spectrograph and was retrieved from the ESO archive. We convolved the UVES spectrum down to the resolution of the MagE and Goodman spectrographs before subtraction. 

\section{Analysis and Results} \label{sec:result}

Using the data in \S \ref{sec:observation}, we carry out detailed modeling to determine mass accretion rates in CVSO~1335 at the different epochs of observations, and estimate its accretion geometry.

\subsection{Accretion Shock Model} \label{ssec:shock}

We use accretion shock models from \citet{calvet1998} to measure the mass accretion rate. This model is based on an assumption that the accretion flow is a cylindrical column in which the material flows vertically onto the stellar surface. At the height where the ram pressure of the flow and the thermodynamic pressure of the stellar photosphere are equal, an accretion shock occurs in which the kinetic energy of the flow is released. Approximately half of the X-rays emitted from the shock are absorbed by the material in the incoming flow, the \emph{pre-shock}, and the rest by the \emph{post-shock} region and the stellar photosphere below the shock. Energy reprocessed by these regions emerges as the shock emission. The input parameters of the model are the stellar mass, radius, and effective temperature, the energy flux of the accretion column, and the filling factor \citep{calvet1998}. In general, the energy flux $F$ determines the spectral slope around the Balmer jump, while the filling factor $f$, the fraction of the stellar surface area covered in the accretion flow, influences the strength of the Balmer jump.

To compare the predictions of the model with the observations, we need to add the shock emission to a photospheric+chromospheric template for the star's spectral type. Since TTS are magnetically active, their chromospheric emission may be significant in the UV region for the low accretors \citep{ingleby2013}. We use the spectrum of the non-accreting T Tauri star RECX~1 \citep{ingleby2013} as the template for the stellar photosphere+chromosphere. We used the low-resolution X-shooter spectra of the star, taken from the ESO Archive, for this analysis. Finally, the model spectra are generated by adding the spectral template, the pre-shock, the post-shock, and the heated photosphere.

We created a grid of models varying the energy flux $F$ between $1.0\times10^{10} - 9.0\times10^{11}$\,ergs\,cm$^{-2}$\,s$^{-1}$ and the filling factor $f$ between 0.05 and 2.05\% to find the best fit to the observed spectra. Since the seeing during the observation was higher than the size of the slit, we expected some uncertainty in the absolute flux level. Without simultaneous photometric measurement during the MagE observation, we assumed that the optical spectra were approximately constant as we did not observe an evidence of optical veiling nor did we expect any such veiling in such a low accretor. Therefore, we adjusted the flux level of all spectra to correspond to the star's SDSS~r' magnitude. With this uncertainty, we assumed a conservative estimate of the flux calibration in the UV part at the 10\% level.

\begin{figure*}[ht!]
\epsscale{1.18}
\plotone{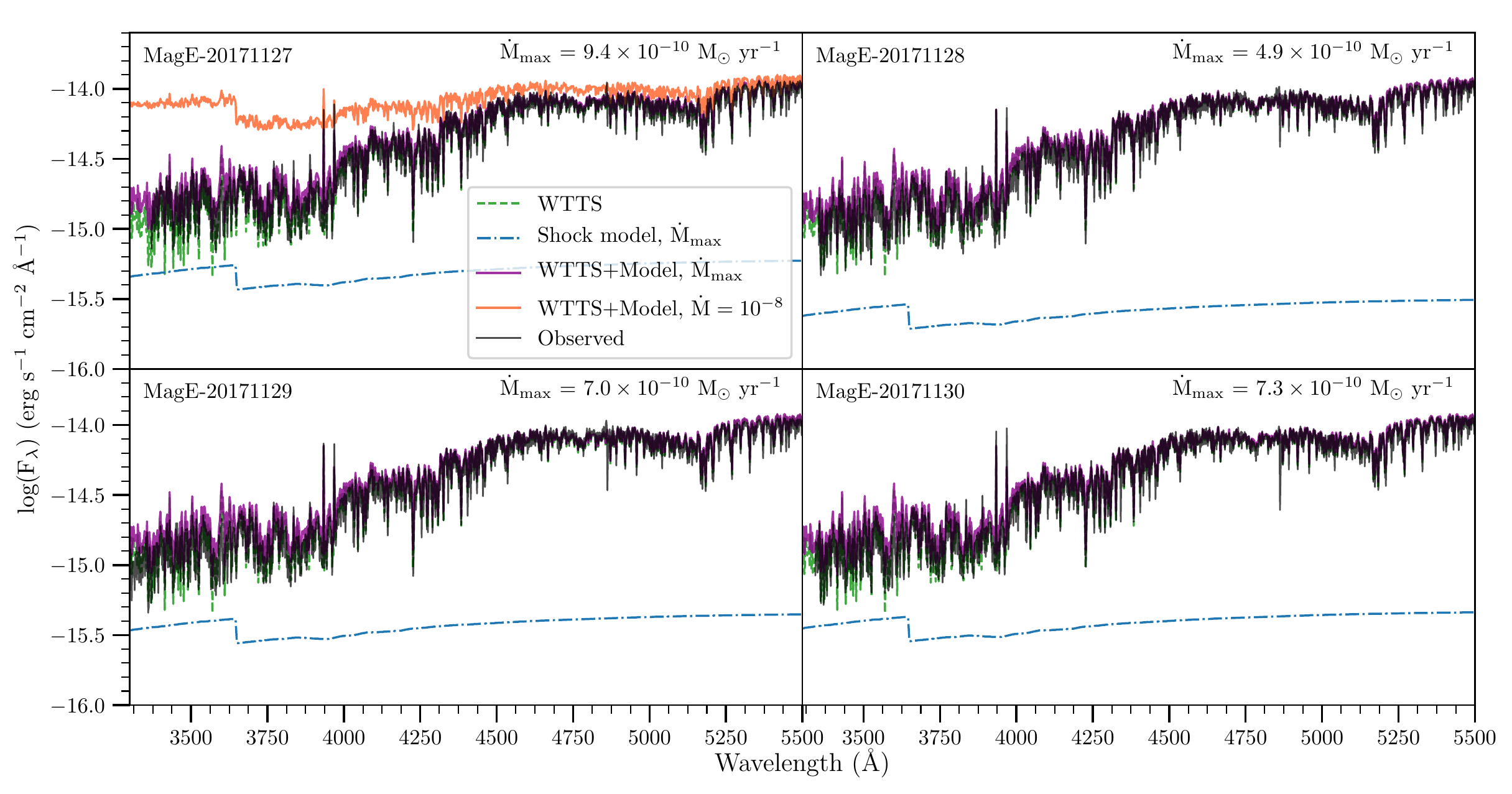}
\caption{The results from the accretion shock model of four MagE spectra of CVSO~1335.
The observed spectra are normalized to the star's SDSS i' magnitude.
No significant excess over the non-accretor standard is detected in all of the spectra. Upper limits are calculated assuming 10\% uncertainty in the flux and photometric calibration. For comparison, the orange line shows the model of the star at the mass accretion rate of $1.0\times10^{-8}\,\msunyr$, a typical accretion rate for T Tauri stars.}
\label{fig:shock}
\end{figure*}

To determine if there is a detection of the Balmer jump due to the accretion shock, in addition to that from the chromosphere, we calculated the integrated flux in the spectral range $3300-5000$\,{\AA} of the observed spectra and compared the measurements with the integrated fluxes of spectra in our grid of shock models.
We found that the predicted fluxes were higher than the observed fluxes in all models, 
but the excess was within the 10\% uncertainty of the flux measurement for a set of models with low accretion rates. Nevertheless, these results suggest that there is no detection of the accretion shock contribution to the Balmer jump in all of our observations and that the jump is mainly chromospheric as is the case in the template WTTS. Therefore, we estimate an upper limit of the mass accretion rate by adopting the highest accretion rate among the models with predicted flux within 10\% of the observed one. Figure~\ref{fig:shock} shows the models with the highest accretion rates of which the fluxes are still consistent with the measurement.
Table \ref{tab:model_shock} shows the upper limits to the mass accretion rates for each epoch. These limits allow for the possibility of accretion variability on a daily timescale.

\begin{deluxetable}{cccc}[t]
\tablecaption{Results of Accretion Shock Models \label{tab:model_shock}}
\tablehead{
\colhead{Obs. Date} & \colhead{\mdot$_{\rm shock}$} & \colhead{log(L$_{\rm acc}$/L$_{\odot}$)} & \colhead{log(L$_{\rm acc}$/L$_{\star}$)} \\
\colhead{UT} & \colhead{($10^{-10}\,\msunyr$)} & \colhead{} & \colhead{}
}
\startdata
20171127    & $\lesssim9.39$ & $\lesssim-1.37$ & $\lesssim-1.19$ \\
20171128    & $\lesssim4.93$ & $\lesssim-1.65$ & $\lesssim-1.47$ \\
20171129    & $\lesssim7.04$ & $\lesssim-1.49$ & $\lesssim-1.31$ \\
20171130    & $\lesssim7.27$ & $\lesssim-1.48$ & $\lesssim-1.30$ \\
\enddata
\end{deluxetable}

\subsection{Magnetospheric Accretion Model} \label{ssec:acc_model}

To estimate the mass accretion rates and the geometry and properties of the accretion flows we modeled the profiles of {\halpha} with magnetospheric accretion models from \citet{muzerolle2001}. The model is described in \cite{hartmann1994,muzerolle1998a,muzerolle2001}, and here we summarize the main assumptions. The model assumes that the geometry of the magnetospheric accretion flows is axisymmetric, following the dipole magnetic field, with the magnetic pole aligned to the stellar and Keplerian disk rotation poles, assumed to be the same. Material flows toward the star inside the boundary controlled by the magnetic dipole geometry and specified by the innermost radius of the disk $R_{i}$ and the width at the base of the flow $W$. The mass flow is steady and set by the total mass accretion rate {\mdot}. The temperature at each point in the flow is a free parameter, and each model is specified by the maximum temperature in the flow $T_{\rm max}$. The models use the extended Sobolev approximation to calculate mean intensities, which in turn are used to calculate radiative rates in the statistical equilibrium for the level populations of a 16-level hydrogen atom \citep{muzerolle2001}. The line flux is determined by using a ray-by-ray method, in which the specific intensity and the total optical depth at each ray are calculated at a given inclination $i$. The final {\halpha} line profile is calculated from the spatially integrated specific intensity.

We calculated a large grid of models with parameters covering ranges shown in Table~\ref{tab:model_param}. The ranges of parameters are chosen to cover all possible values based on previous parametric exploration of \citet{muzerolle2001}. In particular, they found that as {\mdot} decreases, {\tmax} needed to be increased in order to reproduce the observation; {\tmax} between 10k and 12k is required for {\mdot = 10$^{-9}\,\msunyr$}. We adopted this range for our modeling since the expected mass accretion rate is lower than this value, based on the accretion shock model.
To compare the line profile results with the observations, we convolved the model profiles with the instrumental profile of the spectrograph and use the $\chi^2$ statistics to determine the best fit for each observation.

\begin{figure*}[t]
\epsscale{1.1}
\plotone{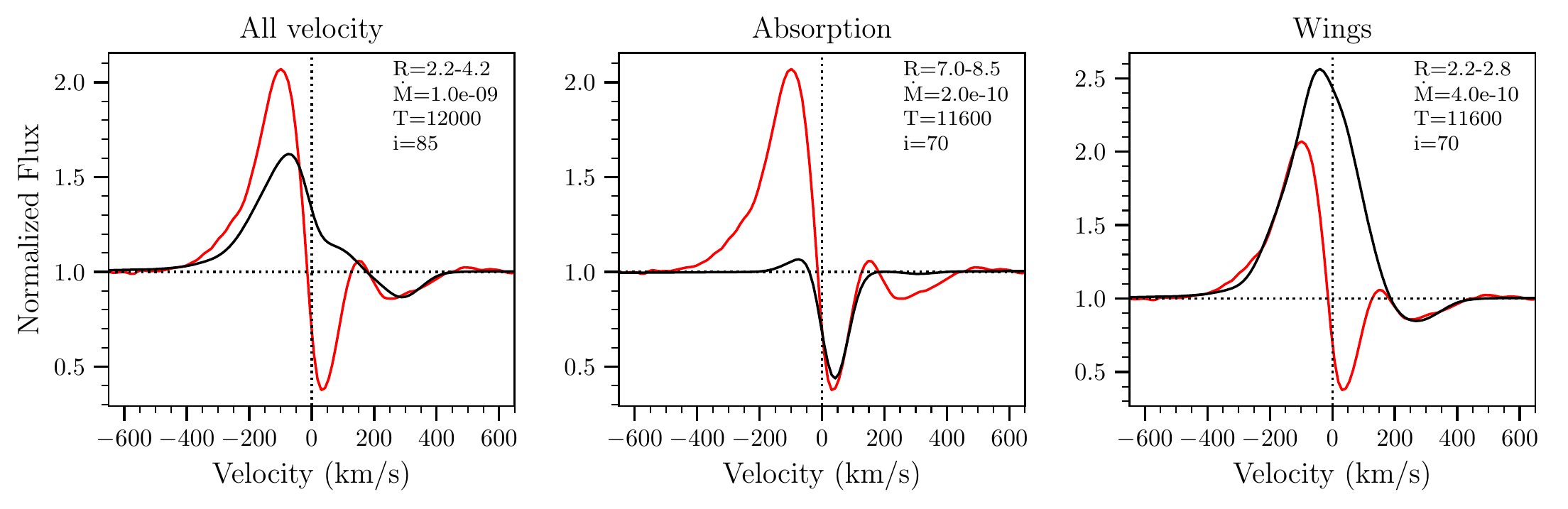}
\caption{Best fits for the {\halpha} line using the standard magnetospheric accretion model. The convolved model line profiles are shown in black, while the observations, taken from 20171130, are shown in red.
\emph{Left:} The best fit for the entire velocity range range. The model could not reproduce the observation.
\emph{Center:} The best fit model for the low-velocity redshifted absorption using a large magnetosphere and a low accretion rate.
\emph{Right:} The best fit model for the entire line profile, excluding the low-velocity redshifted absorption. The profile can be reproduced with a small magnetosphere and a slight increase in the mass accretion rate.}
\label{fig:one_layer}
\end{figure*}

We show examples of fitting line profiles using the standard model in Figure~\ref{fig:one_layer}. First, we attempted to fit the {\halpha} profile for night 20171130 using the entire velocity range of the line. The best fit, shown on the left panel, could not reproduce the the multiple emission and absorption components in the line profiles. Therefore, we attempted to fit each component in the line profile separately.

To fit the low-velocity redshifted absorption component of the line, we selected the model that gave the best fit to observations in the velocity range of 0-150 $\kms$. Qualitatively, we found that models with low mass accretion rates, large magnetospheres, high inclinations, and high temperatures could reproduce the narrow absorption component at $\sim 75 \,\kms$. We then created a grid of models with parameters suitable for exploring these ranges of parameter space. The ranges and the value of the parameters are shown in Table~\ref{tab:model_param}, and an example of the best fit model for these parameters are shown in the center panel of Figure~\ref{fig:one_layer}.

For the rest of the line profile with the standard model, excluding the 0-150 $\kms$ region, we found that the best fit models tended to favor higher accretion rates, small magnetospheres, high inclinations, and high temperatures. We therefore created another set of models to explore these parameters, shown in Table~\ref{tab:model_param}. By excluding the low-velocity redshifted absorption from the fit, the model could fit the wings of the line, including the redshifted absorption component near the free-fall velocity ($\sim$ 200\,$\kms$). Shown in the right panel of Figure~\ref{fig:one_layer} is an example of the best fit for the small magnetosphere.

The general trend from these fits is that the components of the {\halpha} line profile could be reproduced using low mass accretion rate, high temperature, and high inclination. 

\begin{deluxetable}{lccc}[h!]
\tablecaption{Range of Model Parameters \label{tab:model_param}}
\tablehead{
\colhead{Parameters} & \colhead{Min.} & \colhead{Max.} & \colhead{Step} 
}
\startdata
\multicolumn{4}{l}{\emph{Standard Model}} \\
{\mdot} ($10^{-9}\,\msunyr$)                & 0.1   & 9.0   & 0.1, 1.0  \\
$R_{\rm i}$  (R$_{\star}$)                  & 2.2   & 5.4   & 0.4 \\
$W$ (R$_{\star}$)                           & 0.8   & 2.0   & 0.4 \\
$T_{\rm max}$ (K)                           & 10000 & 12000 & 250 \\
$i$ (deg)                                   & 10    & 85    & 5 \\ \hline
\multicolumn{4}{l}{\emph{Small Magnetosphere}} \\
{\mdot$_{\rm in}$} ($10^{-10}\,\msunyr$)    & 1.0   & 9.5   & 0.5  \\
$R_{\rm i, in}$  (R$_{\star}$)              & 1.4   & 3.8   & 0.4 \\
$W_{\rm in}$ (R$_{\star}$)                  & 0.2   & 0.6   & 0.2 \\
$T_{\rm max, in}$ (K)                       & 11000 & 12000 & 200 \\
$i_{\rm in}$ (deg)                          & 55    & 85    & 5 \\ \hline
\multicolumn{4}{l}{\emph{Large Magnetosphere}} \\
{\mdot$_{\rm out}$} ($10^{-10}\,\msunyr$)   & 1.0   & 9.5   & 0.5  \\
$R_{\rm i, out}$     (R$_{\star}$)          & 5.5   & 8.0   & 0.5 \\
$W_{\rm out}$ (R$_{\star}$)                 & \multicolumn{3}{c}{0.3, 0.5, 1.0, 1.5} \\
$T_{\rm max, out}$ (K)                      & 11000 & 12000 & 200 \\
$i_{\rm out}$ (deg)                         & 65    & 85        & 5 \\
\enddata
\end{deluxetable}

\subsection{Modified Magnetospheric Accretion Model} \label{ssec:mod_acc_model}
\subsubsection{Computations with the new geometry} \label{sssec:new_geometry}

\begin{figure}[]
\epsscale{1.15}
\plotone{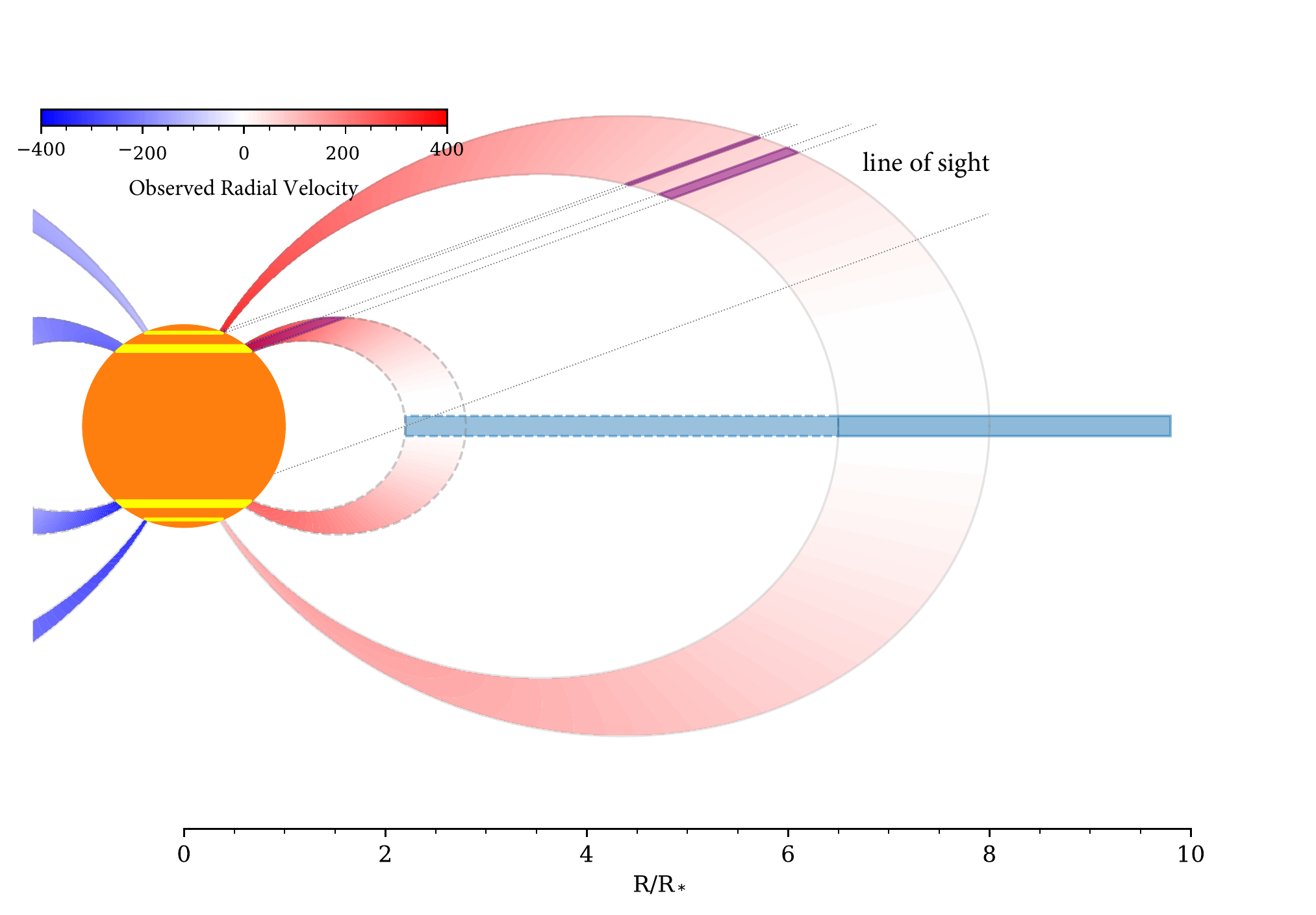}
\caption{The side view schematic of the geometry of the modified magnetospheric accretion model. The material flows from the gas disk onto the star along two geometrically separated, axisymmetric flows. The inner flows originates from $\sim 2.5$\, R$_{\star}$ and the outer flow, covering the entire geometry, originates at $\sim 7$\, R$_{\star}$. The color at each point in the flow represents the relative velocity of the emission from that point as view from an observer at the inclination of 70\degree.}
\label{fig:schematic}
\end{figure}

To fit the entire line profile, we modified the magnetospheric accretion model to include two magnetospheric flows in concentric shells; the \emph{inner flow} resembles a small magnetosphere, and the \emph{outer flow} corresponds to a larger magnetosphere covering the entire accretion structure. A schematic drawing of the new geometry is shown in Figure~\ref{fig:schematic}. We also slightly modify the calculation of the line profile. For a given inclination, the standard model calculates the emerging specific intensity $I_{\nu, p, q}$ and the total optical depth $\tau_{\nu, p, q}$ at each location in the projected coordinate system $(p, q)$ on the sky. To calculate the composite profile with both magnetospheric flows, we assume that the flows are geometrically separated and the source function and level populations of each flows are independent. For the inner flow, the specific intensity includes the emission from the stellar photosphere and the accretion shock, which is absorbed by the accreting material, and the emission from the flow itself. The emission from the outer flow includes only that from the accretion flow. The total emission map of the entire geometry is then given by 
\begin{equation}
\mathcal{I}_{\nu, total} = \exp(-\mathcal{T}_{\nu, outer}) \cdot \mathcal{I}_{\nu, inner} + \mathcal{I}_{\nu, outer}, 
\end{equation} 
where $\mathcal{I}$, $\mathcal{T}$ are 2D maps of the specific intensity and optical depth, respectively. 
Finally, the model line flux is calculated as 
\begin{equation} 
F_{\nu} = \iint \mathcal{I}_{\nu, total} \,dp\,dq. 
\end{equation}

It is computationally impractical to compute the profile from all possible combinations of all small flows and large flows. For each observed profile, we selected 100 best fits for the small and large magnetosphere models, as outlined above. Even though the large magnetosphere could physically be similar in all epochs, the 100 best-fit model profiles of the magnetosphere could be different from epoch to epoch. This is because the depth and shape of the low-velocity redshifted absorption, to which the large magnetosphere models are fitted, still depend on the strength of the emission from the small magnetosphere, especially when that emission is strong. Assuming that the large magnetospheres are similar, which is supported by the persistent nature of the low-velocity component, we produced a global best fit of the large magnetosphere by combining the 100 best fits from each observed model. The final model line profiles for each observed profiles are then calculated by the combination of the 100 best fits of the small magnetosphere and the global best fits of the large magnetosphere.

We selected the best fits in each step using the minimization of $\chi^2$, given by
\begin{equation}
    \chi^2 = \sum_i\, \frac{\left(F_{{\rm obs}, i} - F_{{\rm model}, i}\right)^2}{F_{{\rm obs}, i}},
\end{equation}
where $i$ indexes over the pixels in the observed spectra in relevant velocity ranges. In addition, we have experimented with other fitting methods including Root Mean Square Error (RMSE), the Mean Absolute Deviation (MAD), and the Mean Absolute Percentage Error (MAPE) of each models. The mean and the standard deviation of the first 100 best fits of these statistics are very similar to those using the $\chi^2$ fit, suggesting that the choice of statistical tools does not affect the general results.

\subsubsection{General Results for {\halpha}} \label{sssec:halpha}
\begin{figure*}[t]
\epsscale{1.18}
\plotone{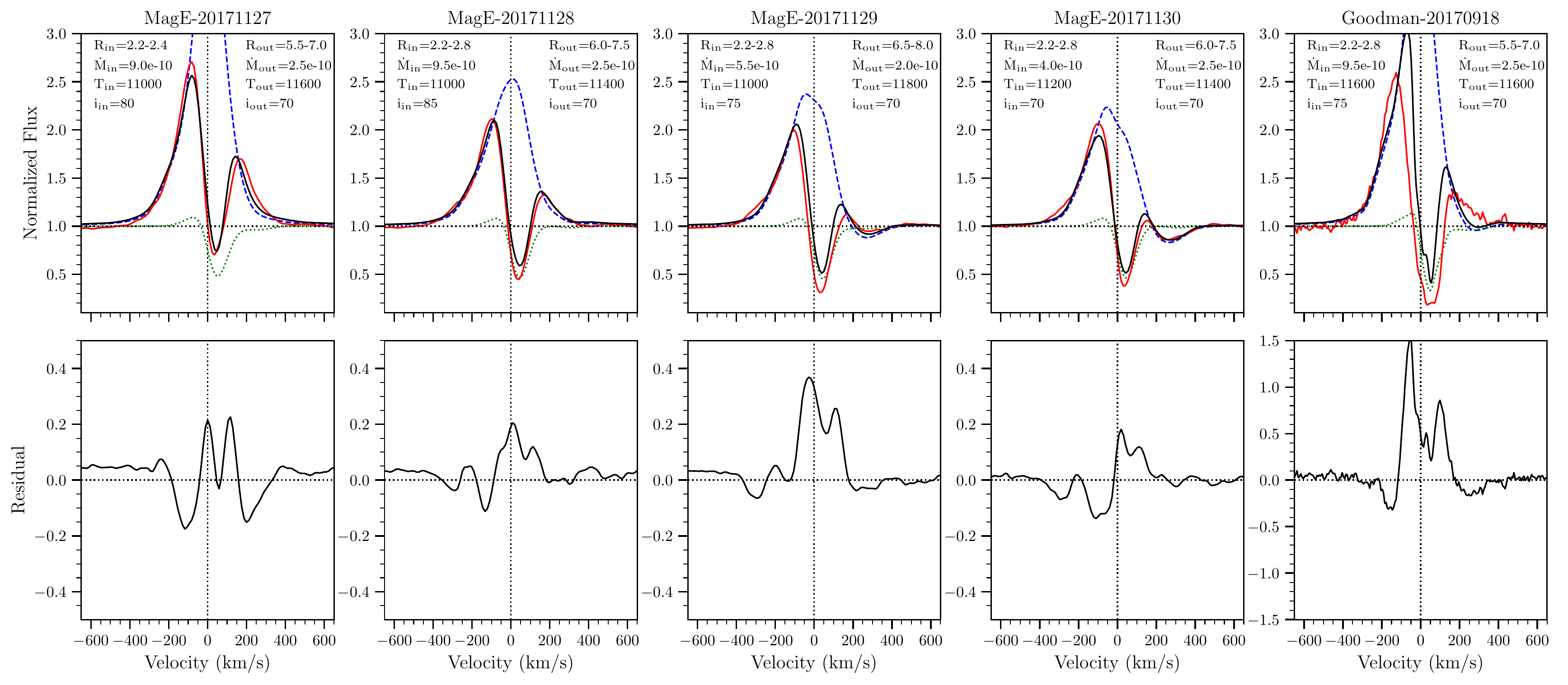}
\caption{The best fits from the modified magnetospheric accretion model for {\halpha}. The top row shows the comparisons between the observed profiles (red) and the convolved model profiles (black). The green dotted lines and the blue dashed lines are the flux from the outer and inner flows, respectively. The bottom row shows the residual between the observed and the model profile (F$_{\rm model}$ $-$ F$_{\rm obs}$). Our model with two accretion flows can qualitatively reproduce the observed line profiles in all epochs, spanning several months. The model has some difficulty fitting the Goodman spectra, possibly due to its higher resolution that the model could not account for. However, the general results are clear that multiple flows are required to fit the profile. Evidently, the model is missing some absorption component at $\sim100\,\kms$, which may indicate even more complex geometry.}
\label{fig:two_geometry}
\end{figure*}

Figure~\ref{fig:two_geometry} shows the $\chi^2$ best fit model for each of the five {\halpha} observations. 
Our two-shell models are able to qualitatively reproduce the observed profiles in all epochs. 
Table \ref{tab:model_results} shows the mean and standard deviation of model parameters from 
the first 1000 $\chi^2$ best fits for each observed profile. These model profiles are qualitatively similar.

For both inner and outer flows the models requires high temperatures, with T$_{\rm max}\sim11000$\,K. The inclination for the outer flow is fairly constant at $\sim70\degree$, while that of the inner flows vary slightly. Similarly, the mass accretion rates for the outer flow are quite steady at $\sim2\times10^{-10}\,\msunyr$, while the accretion rates for the inner flow are somewhat more variable. These results suggest that the inner flow and the outer flow are slightly misaligned, and the outer flows are more stable than the inner flows.

The corotation radius R$_c$, outside which mass cannot accrete onto the star, is an absolute upper limit of the size of the magnetosphere. As a consistency check, we calculate R$_c$ of the star, assuming that the disk plane is aligned with the equatorial plane of the stellar rotation. Since the rotation period of the star is still undetermined, we use the measured projected rotational velocity $v\sin(i)$ of the star as a proxy. In this case, the corotation radius is given by
\begin{equation}
    R_c = \left(GM\right)^{1/3}\left(\frac{R_{\star}\sin(i)}{v\sin(i)}\right)^{2/3},
\end{equation}
where $i$ is the inclination of the system. 
With $v\sin(i)=11.5\,\kms$ based on APOGEE results (J. Hernandez, private communication), we found that R$_c\sim8.9$\,R$_{\star}$ for $i=70\degree$. This is consistent with the model results for the outer flows (Table~\ref{tab:model_results}) and suggests that the infall originates close to the stellar corotation radius.

We calculated the residual of the model from the observation (Figure~\ref{fig:two_geometry}, lower panel), and found that the model systematically over-predicts some emission on the red side of the profile at $\sim 100\,\kms$. This may indicate that the absorption component of the model is not extended enough, suggesting that there is additional absorbing material that is not accounted for.

\subsubsection{Testing the Model with {\hbeta}} \label{ssec:hbeta}

To test the consistency of the two-shell model, we applied the modeling set up to the {\hbeta} lines observed simultaneously with {\halpha}. We found that, similarly to the case of {\halpha}, the one-flow geometry could not reproduce the {\hbeta} observations. We therefore followed the procedure outlined in \S\ref{sssec:new_geometry} to model the {\hbeta} lines.

\begin{figure*}[t]
\epsscale{1.18}
\plotone{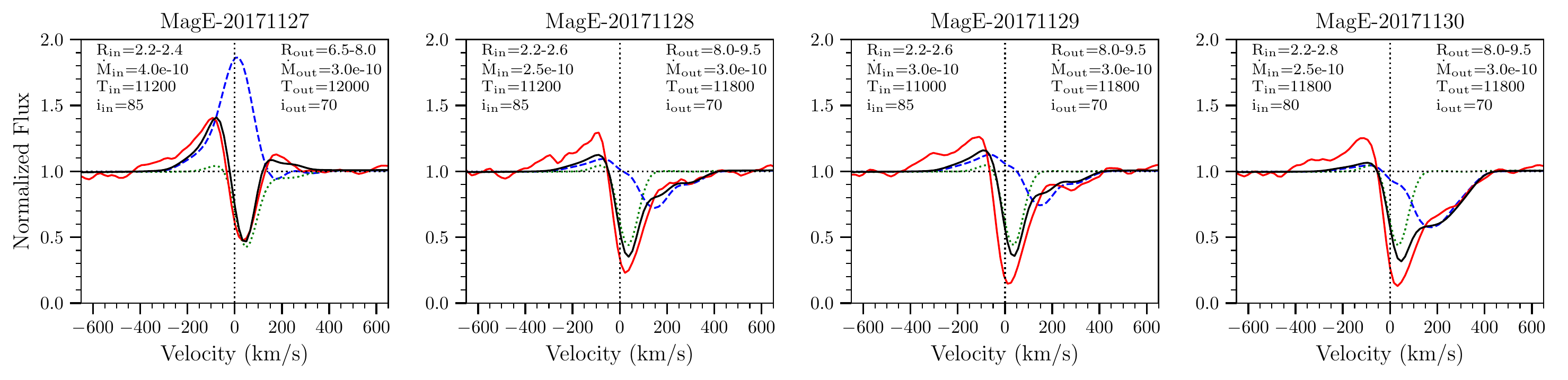}
\caption{The best fits from the modified magnetospheric accretion model for {\hbeta}. The legends are the same as in Figure~\ref{fig:two_geometry}. The modified geometry qualitatively reproduced the observation in all epoch, especially the emission and the high-velocity redshifted absorption. The deviation between the model and the observation at the low-velocity redshifted absorption is more prominent for this line compared to the {\halpha} line models. 
}
\label{fig:hbeta}
\end{figure*}

Figure~\ref{fig:hbeta} shows the best fits of the four {\hbeta} profiles. The models are able to qualitatively reproduce the observation in all epochs. However, the models
cannot entirely fit the strong low-velocity redshifted absorption in three epochs, suggesting that extra absorbing material is needed. As shown in Table~\ref{tab:model_results}, the geometries, temperatures, and inclinations of both the inner flows and the outer flows for {\hbeta} are consistent with those of {\halpha}.
In fact, {\hbeta} profiles calculated using the parameters that produce the best fits for {\halpha} show two redshifted absorption components as seen in the observed {\hbeta} profiles and the best fit profiles. This seems to indicate that the global properties of the flows derived from both lines are similar.
However, in the best-fit models for H$\beta$ the mass accretion rates in the outer flows for {\hbeta} are slightly higher than those for {\halpha}, and the opposite is found for the inner flow. 
In addition, the discrepancy between the accretion rates in the inner flows and outer flows become smaller for {\hbeta}. This could indicate that {\halpha} and {\hbeta} are formed in a slightly different region in the accretion flows with different filling factors, or it could indicate deviations in the temperature distribution assumed in our model.

\subsubsection{Model Limitation}
Stellar winds and disk winds may be present in accreting stars surrounded by protoplanetary disks. In fact, high-sensitivity and high spatial resolution observation have shown the presence of neutral hydrogen emission component from an extended region close to the central accreting stars, which is likely to be from winds \citep{gravity-collaboration2017,koutoulaki2018}. Our magnetospheric accretion model does not include a wind component, but we do not expect a significant contribution of the line emission from winds, since the mass accretion rate is low \citep{muzerolle2001}.

Another limitation of the model is based on the axisymmetric assumption. Simulations \citep{romanova2003,romanova2004} have shown that the accretion flow is generally not axisymmetric and the magnetic pole is likely not aligned to the rotation axis. However, without a measurement of the magnetic properties of CVSO~1335, a parametric study using MHD simulations would be prohibitively expensive. Our model allows a parametric study with a significantly smaller resources requirement.

Lastly, the model assumes only a strictly dipolar geometry, which is likely not the case for more evolved T Tauri stars such as CVSO~1335. However, adding a complex flow prescription would add more parameters to the study and the model would lose its generality. By keeping the geometry simple, we can make simple inferences about the relationship between the parameters in the model.

\begin{deluxetable*}{lcccccccccc}
\tablecaption{Results of the Modified Magnetospheric Accretion Model \label{tab:model_results}}
\tabletypesize{\scriptsize}
\tablehead{
\colhead{Spectrum} &
\colhead{\mdot$_{\rm in}$} &
\colhead{$R_{\rm i, in}$} &
\colhead{$W_{\rm in}$} &
\colhead{$T_{\rm max, in}$} &
\colhead{$i_{\rm in}$} &
\colhead{\mdot$_{\rm out}$} &
\colhead{$R_{\rm i, out}$} &
\colhead{$W_{\rm out}$} &
\colhead{$T_{\rm max, out}$} &
\colhead{$i_{\rm out}$} \\
\colhead{} &
\colhead{$10^{-10}\,\msunyr$} &
\colhead{R$_{\star}$} &
\colhead{R$_{\star}$} &
\colhead{$10^4\,$K} &
\colhead{deg} &
\colhead{$10^{-10}\,\msunyr$} &
\colhead{R$_{\star}$} &
\colhead{R$_{\star}$} &
\colhead{$10^4\,$K} &
\colhead{deg}
}
\startdata
\multicolumn{11}{l}{\halpha} \\
Goodman-20170918  &  8.2$\pm$1.4  &  2.2$\pm$0.0  &  0.4$\pm$0.1  &  1.15$\pm$0.03  &  72$\pm$3  &  2.4$\pm$0.4  &  5.8$\pm$0.5  &  1.5$\pm$0.1  &  1.14$\pm$0.03  &  70$\pm$0 \\
MagE-20171127     &  7.9$\pm$1.7  &  2.2$\pm$0.1  &  0.2$\pm$0.0  &  1.15$\pm$0.04  &  75$\pm$8  &  2.4$\pm$0.4  &  5.9$\pm$0.5  &  1.5$\pm$0.1  &  1.14$\pm$0.03  &  70$\pm$0 \\
MagE-20171128     &  8.8$\pm$0.6  &  2.2$\pm$0.0  &  0.6$\pm$0.1  &  1.15$\pm$0.03  &  85$\pm$1  &  2.3$\pm$0.6  &  6.4$\pm$0.9  &  1.5$\pm$0.1  &  1.14$\pm$0.03  &  71$\pm$2 \\
MagE-20171129     &  5.3$\pm$1.0  &  2.2$\pm$0.0  &  0.6$\pm$0.1  &  1.14$\pm$0.03  &  77$\pm$3  &  2.4$\pm$0.5  &  6.6$\pm$0.9  &  1.5$\pm$0.1  &  1.14$\pm$0.03  &  70$\pm$1 \\
MagE-20171130     &  5.2$\pm$1.9  &  2.1$\pm$0.2  &  0.6$\pm$0.0  &  1.14$\pm$0.03  &  69$\pm$4  &  2.3$\pm$0.4  &  6.3$\pm$0.8  &  1.5$\pm$0.1  &  1.15$\pm$0.03  &  70$\pm$0 \\ \hline
\multicolumn{11}{l}{\hbeta} \\
MagE-20171127     &  3.1$\pm$1.5  &  2.2$\pm$0.0  &  0.2$\pm$0.0  &  1.13$\pm$0.03  &  72$\pm$14  &  3.5$\pm$0.7  &  6.8$\pm$0.8  &  1.5$\pm$0.1  &  1.15$\pm$0.03  &  70$\pm$0 \\
MagE-20171128     &  3.6$\pm$1.2  &  2.2$\pm$0.0  &  0.5$\pm$0.1  &  1.14$\pm$0.03  &  85$\pm$0  &  3.5$\pm$0.7  &  7.1$\pm$0.8  &  1.5$\pm$0.1  &  1.15$\pm$0.03  &  71$\pm$2 \\
MagE-20171129     &  3.9$\pm$1.1  &  2.2$\pm$0.0  &  0.5$\pm$0.1  &  1.14$\pm$0.03  &  85$\pm$0  &  3.5$\pm$0.8  &  7.0$\pm$0.9  &  1.5$\pm$0.1  &  1.15$\pm$0.03  &  71$\pm$2 \\
MagE-20171130     &  3.9$\pm$1.0  &  2.0$\pm$0.2  &  0.6$\pm$0.0  &  1.15$\pm$0.04  &  82$\pm$2  &  3.5$\pm$0.8  &  7.1$\pm$0.9  &  1.5$\pm$0.1  &  1.15$\pm$0.03  &  71$\pm$2 \\
\enddata
\end{deluxetable*}

\subsection{Effects of Inclination and Mass Accretion Rate} \label {ssec:inclination_mdot}
\begin{figure*}[h!]
\epsscale{1.1}
\plotone{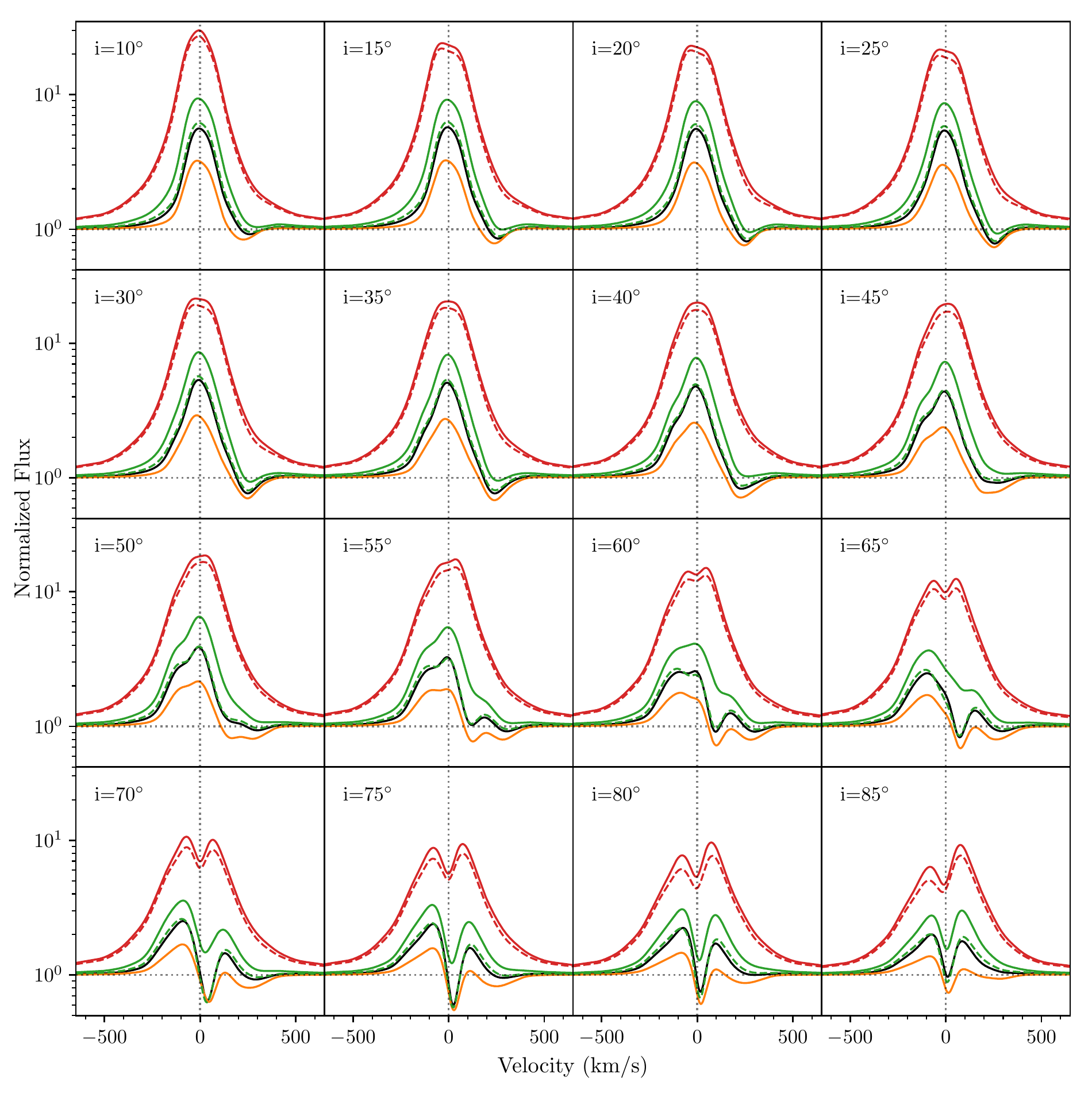}
\caption{
{\halpha} line profiles for the two-shell accretion geometry with varying inclinations, mass accretion rates, and temperatures. Note that the y-axis is in the log scale. The black lines show the {\halpha} profiles with parameters similar to those in Table~\ref{tab:model_results}, with \mdot$_{\rm in}=6\times10^{-10}\,\msunyr$ and \mdot$_{\rm out}=2\times10^{-10}\,\msunyr$. The orange, green, and red lines show the profiles at 0.5, 2.0, and 5.0 times the total mass accretion rate of the black line, respectively. For each {\mdot} the dashed lines show the profiles with a slightly lower T$_{\rm max}$ (see text.)
The low-velocity redshifted absorption starts appearing at $i\geq65\degree$, shifting closer to the line center as the inclination increases. The line profiles at low inclinations are similar to those typically observed.
\label{fig:inclination}}
\end{figure*}

Given that there are some similarities of the {\halpha} and {\hbeta} line profiles of CVSO~1335 to those of dipper stars such as AA~Tau \citep[][see \S\ref{ssec:redshift}]{bouvier2007}, it is insightful to explore the effects that could change the observed line profiles while keeping the two-flow geometry, especially in assessing the frequency of a type of profile. In Figure~\ref{fig:inclination}, we show {\halpha} profiles using the two-flow geometry with parameters similar to the average results in Table~\ref{tab:model_results} but changing the inclination and the total mass accretion rate. Specifically, we selected R$_{\rm i, in}=2.2$\,R$_{\star}$, W$_{\rm in}=0.6$\,R$_{\star}$, with constant ratio of mass accretion rate between the inner and outer flow \mdot$_{\rm in}$/\mdot$_{\rm out}$=3. For the model with high {\mdot}, we also calculated the profiles with a slightly lower temperature ($T_{\rm max, in}$=10000\,K and $T_{\rm max, out}$=11000\,K, compared to 11400\,K and 11600\,K for the fiducial model). We found that the low-velocity redshifted absorption (v$\,\lesssim100\kms$) only appears in models with low total accretion rates in moderate to high inclination, $i\gtrsim65\degree$, in agreement with our findings in \S\ref{ssec:acc_model}. The velocity of the deepest absorption moves closer to the line center as the inclination increases. On the other hand, the high velocity redshifted absorption (v$\gtrsim200\,\kms$) appears in almost all inclinations at low accretion rates, and the velocities of such absorptions are fairly constant. There is some degree of degeneracy between the mass accretion rate and temperature, as models with lower temperature and higher accretion rate are similar to those with higher temperature and lower accretion rate (e.g. green dashed line and black solid line in Figure~\ref{fig:inclination}).

\begin{figure}[h!]
\epsscale{1.1}
\plotone{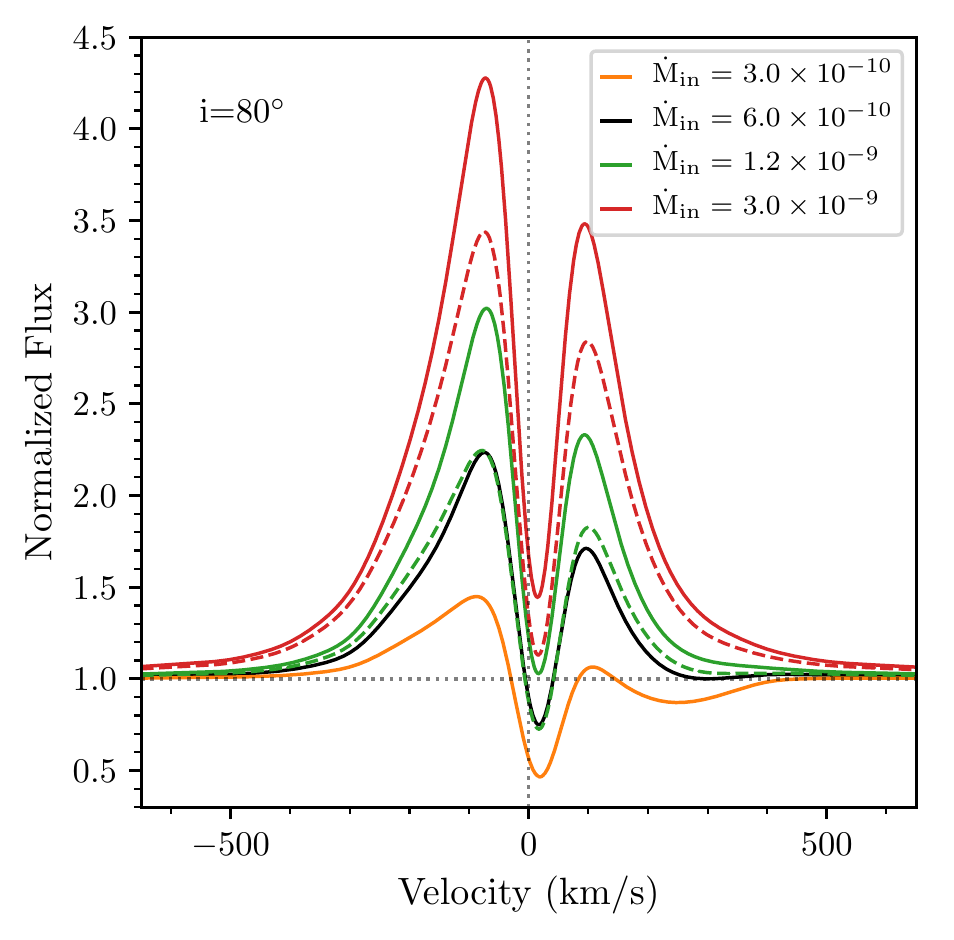}
\caption{Model of {\halpha} line in the same two-shell accretion geometry at different mass accretion rates and temperatures. The legends are the same as in Figure~\ref{fig:inclination}, but the accretion rate of the outer flow {\mdot$_{\rm out}$} is fixed at $2\times10^{-10}\,\msunyr$.
\label{fig:fixout}}
\end{figure}

Figure~\ref{fig:fixout} shows the {\halpha} line profiles with the same parameters as those in Figure~\ref{fig:inclination}, but with a constant \mdot$_{\rm out}=2\times10^{-10}\,\msunyr$ while varying the \mdot$_{\rm in}$ from $3\times10^{-10}$ to $3\times10^{-9}\,\msunyr$.
Interestingly, the low-velocity redshifted absorption appears conspicuously regardless of the mass accretion rate of the inner flow. This suggests that the outer flow could be responsible for a significant amount of emission if the accretion rate is high enough, as in Figure~\ref{fig:inclination}. The outer flow acts as an absorber only in a very low accretion regime ($\sim2-4\times10^{-10}\,\msunyr$).

\section{Discussion} \label{sec:discussion}
\subsection{Measuring Mass Accretion Rates in Low Accretors}
As we have shown in \S \ref{ssec:shock}, the mass accretion rate of CVSO~1335 could be variable, with an upper limit of the order of $\sim 4-9\times10^{-10}\,\msunyr$. These estimates disagree with the mass accretion rate measurements using traditional methods such as the full width at the 10\% height \citep[{\wten};][]{natta2004}. In a previous study \citep{thanathibodee2018}, we calculated the accretion rate of CVSO~1335 using the {\wten} method and found that the accretion rates should be of the order of $10^{-8}-10^{-7}\,\msunyr$ to account for the width of the {\halpha} line. To account for the presence of the redshifted absorption, we estimated {\wten} by measuring the half-width on the blue side, in which no absorption is present. The resulting full width of $\sim600\,\kms$ still gives \mdot$\sim8\times10^{-8}\,\msunyr$. As shown in Figure~\ref{fig:shock}, such an accretion rate would produce a significant excess in the blue part of the optical spectrum of the star, in disagreement with the observation. Our results show that in low accretors such as CVSO~1335, the width at the 10\% height of the {\halpha} line is not a reliable method for measuring the mass accretion rate. The absorption near the line center complicates the measurement of the W$_{10}$ as the actual height cannot be reliably measured.

On the other hand, the mass accretion rates of CVSO~1335 determined in this study by using the magnetospheric accretion model and accretion shock model to fit Balmer line profiles and the Balmer jump are consistent with the accretion rates determined from the {\halpha} luminosity in \citet{thanathibodee2018}, suggesting that {\halpha} line luminosity is a reliable mass accretion rate estimator. Nevertheless, there is a limit to the estimates of mass accretion rates using line luminosities imposed by the chromospheric contribution to the emission lines \citep{manara2013}. The chromospheric contribution is significant in low accretors, since the accretion-originated line emission is weak, and the resulting mass accretion rate would have a high relative uncertainty. On the other hand, line profile modeling can disentangle between the narrow chromospheric core emission and the broad line wings from the accretion flow \citep{espaillat2008}. Therefore, direct modeling of the {\halpha} lines, with chromospheric emission taken into account, is required for an accurate measurement of accretion rates in the low accretion regime and/or with a presence of low-velocity redshifted absorption.

As shown by \citet{ingleby2011b}
emission at the Balmer jump is difficult to detect in low accretors, including the case of CVSO~1335 as shown here. This is because the accretion shock emission is weak compared to the photospheric and chromospheric emission in the UV. Nevertheless, the non-detection of the Balmer jump sets a useful upper limit on the mass accretion rate for objects in which other accretion indicators, such as redshifted absorption, are present.

\subsection{The Origin of the Low-Velocity Redshifted Absorption} \label{ssec:redshift}

The {\halpha} profiles of CVSO~1335 show two systems of redshifted absorption. One is a persistent low-velocity ($v\sim75\,\kms$) redshifted absorption seen in all epochs; another is very variable complex system, located at velocities consistent with free-fall velocities (c.f. Fig.~\ref{fig:observation}.)

Stars classified as dippers, such as AA~Tau \citep{bouvier1999,bouvier2003,bouvier2007} and LkCa~15 \citep{alencar2018}, show comparable redshifted absorptions. Using line profile decomposition, these studies indicate the presence of low-velocity blueshifted and redshifted absorption  components, attributed to a hot wind and to the magnetospheric accretion flow, respectively, in a system viewed at high inclination. The velocities of the blue and redshifted absorption components are correlated and vary  in absolute value between $\sim$ 10 and 60 $\kms$, such that the highest redshifted velocity corresponds to the lowest blueshifted velocity. The low velocity absorption components do not seem to be correlated with the stellar rotation period, unlike the system at nearly free-fall velocities, which is highly variable and tends to appear near the photometric minimum in the dippers; these absorption are also attributed to magnetospheric infall \citep{bouvier1999,bouvier2003,bouvier2007,alencar2018}.

The observed low- and high-velocity redshifted absorptions in the {\halpha} profiles of CVSO~1335 could in principle correspond to those observed in dippers. The star does not seem to exhibit any blueshifted absorption in any of its emission lines, possibly due to its low mass accretion rate, and consequently low mass loss rate. The spectral resolution of our MagE data could not significantly detect temporal variations in the location of the low velocity absorption of the order of those seen in dippers, although the comparison between the MagE and Goodman spectra, spanning over two months, seems to suggest that the absorption are fairly stable within the low velocity range.

Another characteristic of dippers are episodes of dust obscuration observed via reddening of the stellar spectra. In CVSO~1335, the spectrum observed on 2017-11-27 is redder than in other epochs (cf. Figure \ref{fig:observation}), which could be due to a similar reddening process. In support of this possibility, we found that we could reproduce the mean of the spectra in other epochs, adopted as template, by correcting the 2017-11-27 spectrum for extinction using the opacity of silicate dust grains with a size distribution $n(a) \propto a^{-3.5}$ between $a_{min} = 0.005 \mu$m and $a_{max} = 0.25 \mu$m, and a dust-to-gas mass ratio of 0.004, comparable to ISM grains \citep{dalessio2001}. We estimate that a low column density of N$_{\rm gas+dust}\sim9\times10^{-3}$\,g\,cm$^{-2}$ is required to produce the observed obscuration. A more detailed analysis using time-series optical spectra is required to confirm the existence of dust obscuration events in this star.

The high velocity redshifted absorption in CVSO~1335 is highly variable, as it is in dippers, but our limited number of observations does not allow us to determine if it correlates with the stellar period, which we expect to be $\sim 6.5$ days based on a $v\sin(i)$ measurement (\S \ref{sssec:halpha}). We require multi-band photometric monitoring of the star to test the hypothesis that the star is a dipper and observations are under way. Contemporaneous optical-NIR spectroscopic observations are also needed to link the light-curve variation in the photometry to the structure of accretion. Finally, HST observation using high-resolution FUV spectrographs would give insights into the connection between all components in the inner disk and the accretion in this low accretor. These properties will be followed up in future studies.

\subsection{Magnetospheric Accretion in Two-Shell Geometry} \label{ssec:acc_in_two_shell}

The presence of accretion-originated low-velocity redshifted absorption \emph{in addition} to redshifted absorption near free-fall velocity and broad wings suggests that there are two distinct accreting components, which we called two accretion flows in a two-shell geometry (c.f. Fig.~\ref{fig:two_geometry}, Table~\ref{tab:model_results}). This structure is an idealization of the real geometry that must be much more complicated. Depending on the actual structure of the accretion geometry, the difference in mass accretion rate between the inner and outer flows could be explained in different ways. If the inner and outer flow are physically disconnected, the mass accretion rate of the flows would naturally be different. On the other hand, if the two flows represent one complex, but physically connected, accretion flow, the actual mass accretion rate should be the same. The difference in {\mdot} we find could be due to different azimuthal coverage of the inner and outer flows. We note that since the low-velocity redshifted absorption component is stable, the azimuthal coverage of the outer flow must be fairly uniform. In this case, if the inner flow is concentrated in a smaller azimuthal region, the density in the flow would be higher than that in the outer flow which covers a larger volume. Since the model assumes axisymmetric flows, the mass accretion rate inferred from the concentrated rate in the inner flow would be higher than the actual rate.

Although our model could not determine the actual geometry in the accretion flow, the agreement between our simple models and the observations suggests that the two-flow structure must be representing conspicuous features in the real situation that give rise to the observed features. Here we explore the implication of such structure.

\subsubsection{Complex Stellar Magnetic Field Structure} \label{sssec:complex-mag}

The two-shell geometry could be reminiscent of a complex structure of the stellar magnetic field that gives rise to complex accretion flows. For example, it could be that the accreting material is the combination of magnetic dipolar and multipolar fields \citep{long2007,long2012}. Spectropolarimetric observations of accreting T Tauri stars have shown evidences of higher order magnetic fields in stellar magnetosphere; for example \citet{donati2007,donati2011a} showed that both the dipole and octupole components are present in the (K5) T Tauri star V2129~Oph. Based on these results, \citet{romanova2011} used 3D MHD simulations to show that some material could be channelled onto the star via octupole fields at lower latitude in addition to the dipole accretion at higher latitude. CVSO~1335 could exhibit a similar magnetic field structure, and due to its high inclination the dipole flow could occult the higher-order flows at lower latitude. \citet{alencar2012} calculated {\halpha} line profiles based on the results of the simulations, and the low-velocity redshifted absorption was not present in the profiles, while it is conspicous in CVSO~1335. However, this could be due to the combined effect of a lower inclination and higher mass accretion rate in V2129~Oph compared to CVSO~1335 (see \S \ref{ssec:inclination_mdot}).

Another explanation for the two flow geometry is that the accretion in CVSO~1335 is at the interface between stable and unstable accretion in a strictly dipolar magnetic field. Results from 3D MHD simulation show that in this transition region, matter flows along two large funnels and through several small accretion tongues that deposite matter near the stellar surface \citep{romanova2008,kulkarni2009}. If this is the case for CVSO~1335, the outer flows could be the larger funnels that occult the several small tongues, which are the inner flows. To test either of these hypotheses, spectropolarimetric observations and MHD simulations with the exact stellar parameters of CVSO~1335 are required. 

Lastly, the process responsible for the low-velocity redshifted absorption in this star could be similar to that proposed by \citet{bouvier2003}, based on results of numerical simulations \citep{romanova2012,miller1997}, namely an ``inflated'' magnetosphere, resulting from differential rotation between the star and the disk. 
No radiative transfer model of such geometry has been applied to the Balmer lines in AA~Tau yet, although some progress has been made for modeling the low-velocity blueshifted absorption \citep{esau2014}.

\subsubsection{Complex Inner Disk Structure due to a Planetary Companion} \label{sssec:planet}

Another possibility that gives rise to geometrically separated accretion flows is that the inner disk that provides the material is not radially uniform. One example of such condition is that the inner disk, inside the corotation radius, may exhibit a gap structure at $\sim5\,$R$_{\star}$. In this case, very little material at that radius is flowing into the star along the magnetic field lines, thus leaving a gap between two shells of material flowing from the inner ring and from the ring at corotation. Rings and gaps in protoplanetary disks have been observed in larger scales in sub-mm \citep[e.g.][]{alma2015,long2018b} and infrared scattered light \citep[e.g.][]{avenhaus2018}. Hydrodynamic simulations of protoplanetary disks have shown that these rings and gaps could be the consequences of a planet forming in the disk \citep{bae2017}. However, it is unclear that similar process could happen in the inner disk. Nevertheless, if the gap is produced by a planet, it is unlikely that the planet's orbit is stable inside an actively accreting magnetosphere. \citet{bae2017} have shown that a planet could open a secondary gap at $\sim0.5r_p$, where $r_p$ is the orbital radius of the planet, which implies that the planet could be at $\sim10$\,R$_{\star}$ (a$\sim$0.087\,au, P$\sim$10\,d). Since the Shakura-Sunyaev $\alpha$ parameter could be high in the MRI-driven innermost region of the disk \citep{mohanty2018}, and the mass of the gap-opening planet increases with $\alpha$ \citep{bae2017}, the planet, if it exists, should be massive (several M$_{\rm Jup}$).

Planets have been found to orbit the central stars with orbits of a few days, i.e. hot Jupiters. Many studies have indicated the presence of close-in planets around (non-accreting) weak TTS \citep[e.g.][]{van-eyken2012,mann2016,david2017}, but evidence of such planets around an actively accreting pre-main sequence stars are still sparse. So far, the only proposed planets in such category are a 11.3\,M$_{\rm Jup}$ planet with $\sim$\,9 days period orbiting the 2\,Myr star CI Tau \citep{johns-krull2016}, and a M$_p\sin i\sim19.3$\,M$_{\rm Jup}$ planet with a 24.8 day period around the 0.5\,Myr star AS~205A \citep{almeida2017}, based on the radial velocity method. With the parameters of CVSO~1335, a comparable planet would cause an RV signature of $\sim1\,\kms$; spectroscopic monitoring of the star is required to test this hypothesis.

\subsubsection{Observability of Low-Velocity Redshifted Component}
We have shown that the modified magnetospheric accretion model with a
two-flow geometry is able to reproduce the observed {\halpha} and {\hbeta} profiles of CVSO~1335. Therefore, it is insightful to consider this model for other T Tauri stars. As shown in Figure~\ref{fig:inclination}, producing two redshifted absorption components requires low accretion rates and high inclinations; 
for high accretion rates, the two-flow model produces similar profiles to the standard model since {\halpha} has become optically thick. Nevertheless, the model could be used for other lines that are more optically thin even with high accretion rate, but this remains to be calculated.

For low mass accretion rates the presence of the low velocity component still depends on the inclination of the system (c.f. Figure~\ref{fig:inclination}). Assuming that all stars orient randomly, one would expect that a significant portion of the low accretors exhibit a similar type of profiles. Several studies have shown that complex {\halpha} line profiles are not uncommon \citep[e.g.][]{reipurth1996,antoniucci2017}, but {\halpha} profiles with two redshifted absorption components, as seen in CVSO~1335, are rare even in CTTS with low accretion rates. This suggests that other factors contribute to the formation of such geometry. 
The two scenarios discussed in this section could explain the rarity of multiple redshifted absorption profile.

In order to have complex magnetic fields, required for scenario discussed in \S\ref{sssec:complex-mag}, pre-main sequence stars need to have solar or higher mass and/or old age \citep[e.g.][]{villebrun2019}. In this regard, solar mass low accretors are rare since stars in this mass range account for only $\lesssim10\%$ of a young population \citep[e.g.][]{briceno2019}, and the number of accretors decreases sharply with age. The frequency of low accretors are not yet available, but the number should be less than the frequency of all CTTS at 5\,Myr, $\sim$ 15\% \citep{briceno2019,fedele2010}. Therefore, the upper limit of the frequency of solar-mass low accretor is $\sim$1.5\% of all T Tauri stars in a given population, and $\sim$0.75\% would have a high inclination. This fraction means that only a few stars in a given population could show line profiles similar to those of CVSO~1335, since the number of T Tauri stars in a given population is on the order of a few hundreds \citep[e.g.][]{luhman2018a,luhman2018b} to thousands \citep[e.g.][]{sung2009,briceno2019}. It is therefore conceivable that many solar mass stars could have a magnetosphere in the two-flow geometry at some point in their life, but the frequency of observing them is low and only CVSO~1335 has been identified as such so far.

In the gap opened by a planet scenario, the required mass of the planet is several Jupiter mass. If the planet exist, it would be classified as a hot Jupiter due to its close orbit. Since, only $\sim$ 1\% of solar mass stars, and even less in M-type stars, host such planets \citep{dawson2018}, the expected number of CTTS with CVSO~1335-type profiles is as equally small as in the case of complex magnetic fields.

\section{Summary} \label{sec:summary}

We applied accretion shock models and magnetospheric accretion models to MagE and Goodman optical spectra of the 5\,Myr old, $\sim$ solar mass star CVSO~1335 to characterize the accretion properties of this low accretor. Here we summarize our findings:

\begin{enumerate}
    \item The Balmer jump of CVSO~1335 does not show any significant excess over the WTTS used as template, confirming that CVSO~1335 is a low accretor. Using accretion shock models, we find variable upper limits of the mass accretion rates $\sim4-9\times10^{-10}\,\msunyr$. These limits are in agreement with estimates based on {\halpha} line luminosity. However, they contradict the measurements based on the line width, which would indicate rates as high as $10^{-8}\,\msunyr$. 
    
    \item The excess at the Balmer jump does not provide an
    estimate of the mass accretion rate in CVSO~1335, or in general in low accretors. On the other hand, line profile fitting provides a 
    measurement of the accretion rate, as well as the geometry of 
    the accretion flows. Therefore, modeling line profiles is the only reliable method to accurately measure mass accretion rate in low accretors.
    
    \item Redshifted absorption components superimposed on 
    the {\halpha} and {\hbeta} emission lines are conspicuous in CVSO~1335
    and are found in all epochs. The line profiles are variable with clear multiple components, which cannot be explained using the standard magnetospheric accretion model. A modified magnetospheric accretion model, with two separated accretion flows, can explain the low-velocity redshifted absorption simultaneously with broad wings and high-velocity redshifted absorption in the {\halpha} and {\hbeta} lines. The inner flows have higher and more variable {\mdot} compared to the outer flows. High inclination ($i\sim70\degree$) and high accretion flow temperature (T$_{\rm max} \gtrsim 11000$) are required to reproduce the profiles.
    
    \item The required high inclination, in addition to the presence of a persistent low velocity component and a highly variable high velocity component, with velocities
    comparable to free-fall velocities, may indicate that CVSO~1335 is a dipper. Multi-band photometric monitoring is required to test this hypothesis.
    
    \item Our simple two-flow geometry represents a more complex accretion geometry. This could suggests 
    a complex magnetic field structure including higher-order fields, unstable accretion, or an inflated magnetosphere. Alternatively, a ringed structure in the inner disk resulting from the presence of a companion, could be responsible for this geometry. The rarity of {\halpha} profiles with multiple redshifted absorption components is compatible with both scenarios.
\end{enumerate}

\acknowledgments
\noindent
This project is supported in part by NASA grant NNX17AE57G. This research is based on data obtained from the ESO Science Archive Facility under request number thanathi/412978. We thank Lee Hartmann, Jaehan Bae,  Zhaohuan Zhu, Susan Edwards, and Jerome Bouvier for insightful suggestions.
We thank the telescope operators and the staff at Las Campanas Observatory for the help during the MagE observations. 

\facilities{Magellan:Baade (MagE), SOAR (Goodman Spectrograph)}

\bibliography{thanathibodee2019}{}
\bibliographystyle{aasjournal}

\end{document}